\documentstyle[prl,aps,multicol,epsf]{revtex}
\begin{document}
\draft

\input amssym.def
\input amssym.tex
\begin{multicols}{2}

{\bf Hecker et al. reply:} In their comment\cite{HW} den Hartog and
van Wees (HW) raise objections against our analysis of the
experimental data presented in\cite{HHAF}. According to HW, we did not
account for the quantum phase incoherence introduced by the
Niobium compounds of the investigated Nb/Au hybrid samples.  If so,
the experimentally derived ratio $X={\rm rms}(G_{\rm NS})/{\rm
rms}(G_{\rm N})$ would differ substantially from the value $X\simeq
2.8$\cite{HHAF}, thereby invalidating the reported agreement between
theory and experiment. Here we show by means of a resistor network
analysis and by reviewing the information {\it already provided
in\cite{HHAF}} that these objections are not justified.

A schematic scetch of the system we are analysing is shown in
Fig.1, where the dark/light shaded region represents the
Au/Nb compound consisting of $N_{\rm Au}/N_{\rm Nb}$ phase
coherent sub-volumes of size $L_{\varphi,\rm Au}/L_{\varphi,\rm Nb}$ and the
black bar represents the interface region between Au and Nb
($L_{\varphi,\rm Au}/L_{\varphi,\rm Nb}$: phase coherence length of
Au/Nb). 
\narrowtext
\begin{figure}
\epsfxsize=8cm
\centerline{
\epsfbox{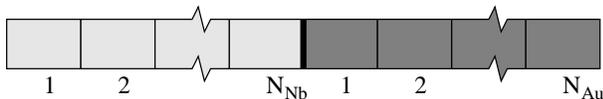}}
\label{fig1}
\caption{Decomposition of the system into incoherent partial resistors
  (see text).}
\end{figure}
The two basic assumptions of the resistor network analysis (cf. Ref.[2]
of\cite{HHAF}) of conductance fluctuations (CF) are that i) the Au/Nb
compound decomposes into $N_{\rm Au}/N_{\rm NB}$ {\it statistically
independent} sub-volumes and ii) that the CF of any of these volumes
are given by the universal value ${\rm rms}(G^0_{\rm NS})$ or ${\rm
rms}(G^0_{\rm N})$, depending on whether the sample is NS or N. An
application of these concepts to the particular system shown in Fig.1
yields the following equations:
\begin{eqnarray}
&&f_{\rm NS}\equiv\frac{{\rm rms}(G^{\rm exp}_{\rm
NS})}{  {\rm rms}(G^0_{\rm NS})}=\frac{1}{(2N_{\rm Au})^{3/2}}\nonumber \\
&&f_{\rm N}\equiv \frac{{\rm rms}(G^{\rm exp}_{\rm
N})}{  {\rm rms}(G^0_{\rm N})}=\sqrt{\frac{1}{N_{\rm Au}^3}
\frac{R_{\rm Au}^4}{(R_{\rm Au} + R_{\rm Nb})^4}}
\label{formulas}
\end{eqnarray}
relating the fully phase coherent to the measured values of the CF,
respectively. (Note that the extra factor of two in 
$f_{\rm NS}$  accounts for the fact that particles contributing to
the NS-conductance have to traverse the Au region twice.)
In (\ref{formulas}) we have set $L_{\varphi,{\rm Nb}}=0$ as the most
conservative estimate (leading to the smallest possible value of
$f_{\rm N}$). Now, the precise value of $N_{\rm Au}$ is unknown to
us, i.e. the best we can do is to fix this parameter by means of the
{\it experimentally} inferred value of $f_{\rm NS}$. In a second
step we deduce a prediction for ${\rm rms}(G^0_{\rm N})$ by means of
the experimental ${\rm rms}(G^{\rm exp}_{\rm N})$, $N_{\rm Au}$ and
the second of the above equations. In this way we find for sample 1
$N_{\rm Au,1}=1.38 \pm 0.12$\cite{fn1} and $X_1=4.2 \pm 0.5$ whilst
for sample 2 $N_{\rm Au,2}=1.50 \pm 0.14$\cite{fn1} and $X_2=3.7 \pm
0.6$. Note that these considerations a) {\it do} account
for the Nb-phase incoherence and b) lead to ratios between NS- and
N-conductance fluctuations which are even larger (roughly by 40 \%) than those
reported in\cite{HHAF}.

How does this analysis relate to what was written in\cite{HHAF}? In
the Letter the effects of incoherence were discussed in terms of two
geometrically estimated effective length parameters, $L^{\rm NS}_{\rm
  eff}$ and $L^{\rm N}_{\rm eff}$, and a global dephasing length
$L_\varphi$. The analysis was less accurate than the one above in that
both the inhomogeneous distribution of partial resistances and the
inequality of $L_{\varphi, {\rm Au}}$ and $L_{\varphi, {\rm Nb}}$ were
neglected. This explains the 40 \% difference between the two results.
Nonetheless {\it the Nb-decoherence was cleary accounted for
  in\cite{HHAF}}. (Otherwise we would have set $L^{\rm N}_{\rm
  eff}=L_{\rm Au}$.) In other words, an application of HW's
'correction' factor to our result would amount to counting the role of
the Nb {\it twice}.

Finally let us point out that the above network analysis can yield no
more than rough estimates of the decoherence factors, too. The point
is that the actual experimental setup is more complex than what
Eqs.(\ref{formulas}) describe: Firstly, the sample production
inevitably leads to the formation of a surface resistance of unknown
size between S and N; secondly, the actual NS-interface is not local
in space but rather extends over a region of 500~nm. The
experimentally measured parameter s do not suffice to unambigiously
fix a resistor model describing this more complicated situation.
However in all realistic trial scenarios we have been analysing in
terms of suitable generalizations of Eqs.(\ref{formulas}), the ratio
between NS- and N-fluctuations was comparable to, if not {\it larger}
than the theoretical prediction.

Summarizing we have attempted to clarify our treatment of phase
decoherence in\cite{HHAF}: Our work did account for both the
incoherence introduced by Au {\it and} Nb and we reject the critisim
as it is formulated in\cite{HW}.  Admittedly, however, the
experimental data allows for no more than an approximate
determination of the decoherence factors (as, indeed, was already
pointed out in \cite{HHAF}). Still we maintain that a massive
enhancement of the NS-fluctuations, comparable with the theoretical
prediction, has been experimentally observed in\cite{HHAF}.

Klaus Hecker$^{1)}$, Helmut Hegger$^{1)}$, Alexander
Altland$^{2)}$, and Klaus Fiegle$^{3)}$\\[0.3cm]
1) II.Physikalisches Institut, Z\"ulpicher Str. 77, D-50937
K\"oln, Germany\\
2) Institut f{\"u}r Theoretische Physik, Z{\"u}lpicher Str. 77, D-50937
K{\"o}ln, Germany\\
3) I.Physikalisches Institut, Z\"ulpicher Str. 77, D-50937
K\"oln, Germany\\

\end{multicols}

\begin{thebibliography}{9}
\bibitem{fn1} Note that these values of $N_{\rm Au}$ are compatible
with previous independent estimates (cf. \protect\cite{HHAF}) for
$L_{\varphi, {\rm Au}}$.
\end{thebibliography}
\end{document}